\documentclass[showpacs,showkeys,preprint,amsmath,amssymb,superscriptaddress]{revtex4}   
\usepackage{graphicx}
\usepackage{dcolumn}
\usepackage{bm}
\newcommand{\eq}{\end{quote}}
\newcommand{\nn}{\nonumber}
\newcommand{\Slash}[1]{\ooalign{\hfil/\hfil\crcr$#1$}}
\newcommand{\rsz}{\resizebox{8cm}{6cm}}
\usepackage{bm}
\begin{document}      
\preprint{PNU-NTG-04/2003}
\title{Regulariztion dependence of $S=0$ and $S=-1$ meson-baryon
system in the chiral unitary model}
\author{S.I.Nam}
\email{sinam@rcnp.osaka-u.ac.jp}
\affiliation{Research Center for Nuclear Physics (RCNP),
Ibaraki, Osaka 567-0047, Japan}
\affiliation{Department of
Physics, Pusan National University, Busan 609-735, Korea}
\author{H.-Ch.Kim}
\email{hchkim@pusan.ac.kr}
\affiliation{Department of
Physics, Pusan National University, Busan 609-735, Korea}
\author{T.Hyodo}
\email{hyodo@rcnp.osaka-u.ac.jp}
\affiliation{Research Center for Nuclear Physics (RCNP), Ibaraki, Osaka
567-0047, Japan}
\author{D.Jido}
\email{jido@ect.it}
\affiliation{ECT*, Villa Tambosi, Strada delle Tabrelle 286, I-38050
Villazzano (Trento), Italy}
\author{A.Hosaka}
\email{hosaka@rcnp.osaka-u.ac.jp}
\affiliation{Research Center for Nuclear Physics (RCNP), Ibaraki, Osaka
567-0047, Japan}

\date{\today}
\begin{abstract}
We investigate the dependence of ${\rm s}$-wave meson-baryon scattering
amplitudes on different regularizations within the framework of the
chiral unitary model.  We employ two different regularization schemes,
{\em i.e.} dimensional and form-factor regulariaztions 
to tame the divergences in the model.  We also study the analytic
structures of $T$-matrices, using those regularization 
schemes.  We find that while the form-factor regularizaion produces
almost the same results as the dimensional regularization did,
the on-shell approximation is to some extent limited in the case of
the form-factor regularization.  Having chosen parameters properly, we
show that the regularization dependences can be minimized.     
\end{abstract}

\pacs{13.75.Gx,13.75Jz,13.85Fb,14.20.Gk}
           
\keywords{Meson-baryon scattering, the chiral unitary model, regularization}
      
\maketitle
\section{Introduction}
Understanding meson-baryon scattering has been a very important issue
for several decades, since it gives information not only on the strong
interaction between hadrons but also on the origin of baryonic
resonances.  Recently, chiral perturbation theory has been of great success in
explaining low-energy meson-baryon scattering, in particular, strangeness ${\rm
S}=0,\, -1$ channels.  While $S$-wave $\pi N$ and $K^+ N$ scattering
can be well described by the Lagrangian of leading order in the  
chiral expansion, the  
$\bar{K} N$ system requires multi-coupled channels in order to generate
resonances such as $\Lambda (1405)$
~\cite{Brown:yv,Lee:1994my}. Though chiral perturbation 
theory (ChPT) provides us with a theoretical and systematic framework to
study meson-baryon systems, it is restricted to lower energy
regimes.  In order to explain the $\bar{K}N$ system quantitatively,
one has to go beyond ChPT.    

Recently, there have been noticeable works on the $\bar{K} N$ systems, 
based on the effective chiral Lagrangian: Kaiser {\em et
al.}~\cite{Kaiser:1995eg} examined the $S$-wave $\bar{K}N$ system.
They utilized three different types 
of the pseudo-potentials, keeping the low energy constants from the
effective chiral Lagrangian and solved the Lippmann-Schwinger (LS)
equation in the coupled-channel formalism.  Krippa and
Londergen~\cite{Krippa:1998ix} investigated also $\bar{K}N$ scattering,
using the on-shell approximation.  Though
Ref.~\cite{Krippa:1998ix} seems to describe the experimental  
data~\cite{Nowak,Tovee} well, threshold behavior can not be reproduced
well in the $\Sigma^-\pi^+$, $\Sigma^+\pi^-$, and $\Lambda\pi^0$
channels due to the openings of new channels.  The Valencia group also studied 
extensively the $s$-wave $\bar{K}N$ system~\cite{Oset:1997it}.
Ref.~\cite{Oset:1997it} started from the effective
chiral Lagrangian to construct the pseudo-potentials and used the
on-shell factorization to solve the coupled-channel LS equation,
employing a regularization with the  three-dimensional cut-off.
Later,~\cite{Oller:2000fj} introduced the dimensional
regularization instead of the three dimensional cut-off.  They also
made use of the N/D method to solve the scattering  equation.  Lutz and 
Kolomeitsev~\cite{Lutz:2001yb} investigated 
meson-baryon scattering in the context of the effective chiral
Lagrangian with the large $N_c$ counting.  They introduced a minimal
chiral subtraction scheme within the dimensional regularization in order
to keep the covariant chiral counting rules.  Though there exist technical 
differences between these works, almost all these works showed a remarkable 
agreement with experimental data.  Thus, it is of great significance to
understand some essence, if any, in dynamics of different technical methods, 
in particular, different regularization schemes and the role of
corresponding parameters.       

In the present work, we turn our attention to the dependence of the
$s$-wave $\bar{K}N$ and $\pi N$ systems on different regularization
schemes.  While the dimensional regularization is preferably selected
in ChPT, the conventional meson-exchange
model~\cite{Machleidt:hj,Mueller-Groeling:cw} gets used to introduce    
form factors, which describe the extended hadron structure, 
as a regularization.  Though the dimensional regularization gains
an advantage over the form factors in many aspects, in particular, in
the context of renormalization and gauge invariance, it is less
convenient to associate with a LS-type scattering equation.  Hence, in
this work, we will show that the use of form factors presents almost 
the same results as that of the dimensional regularization.  The
present investigation will shed light on the meaning of the parameters  
involved in the regularization and will pave the way for solving the
LS scattering equation without on-shell approximation in the future
work by employing the form factors in place of the
dimensional regularization.  

The paper is organized as follows: In section II, we shall explain the
formalism of the chiral unitary model to investigate the meson-baryon
systems.  The analytic structure of the scattering amplitudes will be
presented.  In section III, we shall give the numerical results
in the ${\rm S}=0$ channel with two different regularizations used.  In
section IV, we shall present those in the ${\rm S}=-1$ channel.  In section
V, we shall summarize the present work and draw conclusion with
outlook.    
\section{Formalism}
We start with the effective chiral Lagrangian to the lowest order 
$\mathcal{L}^{1}_{MB}$~\cite{Meissner:1993ah}:
\begin{eqnarray}
\mathcal{L}^{1}_{MB}&=&\langle\bar{B}(i\Slash{\nabla}-M)B\rangle+
\frac{1}{2}D\langle\bar{B}\gamma^{\mu}\gamma_{5}\{u_{\mu},B\}\rangle\nn\\
&+&\frac{1}{2}F\langle\bar{B}\gamma^{\mu}\gamma_{5}[u_{\mu},B]\rangle,   
\label{l1mb}   
\end{eqnarray}
where 
\begin{eqnarray}
{\nabla}_{\mu}B &=&\partial_{\mu} B +[\Gamma_{\mu},B],\nn\\
\Gamma_{\mu} &=&\frac{1}{2}(u^{+}\Slash{\partial}u)+(u\Slash{\partial}u^{+}),\nn\\
U&=&u^{2}=\exp(i\frac{\sqrt{2}\Phi}{f}),\nn\\
u_{\mu}&=&iu^{+}\Slash{\partial}Uu^{+}.
\end{eqnarray}
$B$ and $\Phi$ represent baryon and pseudoscalar fields, respectively:
\begin{eqnarray}
B&=&\left(\begin{array}{ccc}
\frac{1}{\sqrt{2}}\Sigma^{0}+\frac{1}{\sqrt{6}}\Lambda & \Sigma^{+} & p\\
\Sigma^{-} & -\frac{1}{\sqrt{2}}\Sigma^{0}+\frac{1}{\sqrt{6}}\Lambda & n\\
\Xi^{-} & \Xi^{0} & -\frac{2}{\sqrt{6}}\Lambda
  \end{array}\right),\cr
\Phi&=&\left( \begin{array}{ccc}
\frac{1}{\sqrt{2}}\pi^{0}+\frac{1}{\sqrt{6}}\eta & \pi^{+} & K^{+} \\
\pi^{-} & -\frac{1}{\sqrt{2}}\pi^{0}+\frac{1}{\sqrt{6}}\eta & K^{0}\\
K^{-} & \bar{K}^{0} & -\frac{2}{\sqrt{6}}\eta
  \end{array}\right).
\end{eqnarray}
The coefficients $F$ and $D$ denote the reduced matrix elements for
semiletonic decays of octet baryons in SU(3) flavor symmetry.  $f$ in
the $U$ field is the meson decay constant~\cite{Gasser:1984gg},  
and $\langle\cdots\rangle$ denotes the trace over $SU(3)$ flavor
space.  Since we are interested in $s$-wave scattering at low
energies, we need only the 
terms of order ${\cal O}(p)$ in the expansion of the $U$ field:    
\begin{eqnarray}
&&\mathcal{L}^{1}_{MB}=\frac{i}{4f^{2}}\langle\bar{B}\gamma^{\mu}[
(\Phi\partial_{\mu}\Phi-\partial_{\mu}\Phi\Phi)B-B(
\Phi\partial_{\mu}\Phi-\partial_{\mu}\Phi\Phi)]\rangle. 
\label{wt}  
\end{eqnarray}
The Lagrangian in Eq.~(\ref{wt}) is also known as the venerable
Weinberg-Tomozawa term.  The corresponding pseudo-potentials can be
easily obtained as:
\begin{eqnarray}
V_{ij}&=&-\frac{C_{ij}}{4f^{2}}\bar{u}(p_{j})\gamma_{\nu}u(p_{i})
(k_{i}^{\nu}+k_{j}^{\nu}),  
\label{pp}
\end{eqnarray}
where $p_{i}$ and $p_{j}$ ($k_{i}$ and $k_{j}$) are initial and final baryon
(meson) momenta, respectively. The subscripts $i$ and $j$ denote the
indices representing the coupled-channel states.  The coefficients
$C_{ij}$ are derived from Eq.~(\ref{wt}).  Explicit forms of the 
$C_{ij}$ for ${\rm S}=-1$ and ${\rm S}=0$ are given in
Refs.~\cite{Oset:1997it,Inoue:2001ip}.       
 
The next step is to solve the Bethe-Salpeter(BS) equation. The general
$T$ matrix 
in the coupled-channel formalism is given by       
\begin{equation}
T_{ij}=V_{ij}+\sum_{l}i\int\frac{d^{4}q}{{(2\pi)}^{4}}
\frac{V_{il}(\Slash{p}_{l}+M_{l})T_{lj}}{\{{(P-q)}^{2}-M_{l}^{2}\}(q^{2}-m_{l}^{2})},
\label{bs}  
\end{equation}
where $M_l$ and $m_l$ represent the octet baryon and meson masses in the
intermediate state, respectively.  $P$ designates the total momentum
given by $p+q=(\sqrt{s},0,0,0)$ in the center of mass system, where $p$ and $q$ denote
the intermediate baryon and meson momenta, respectively.  Having
utilized the on-mass-shell factorization or N/D method, Eq.~(\ref{bs})
becomes an analytically solvable and 
pure algebraic equation:  
\begin{equation}
T_{ij}=V_{ij}+\sum_{l}V_{il}\left\{i\int\frac{d^{4}q}{{(2\pi)}^{4}}
\frac{2M_{l}}{\{{(P-q)}^{2}-M_{l}^{2}\}(q^{2}-m_{l}^{2})}\right\}T_{lj}.
\label{ombs}  
\end{equation}
The pseudo-potential in Eq.~(\ref{ombs}) is then written by   
\begin{eqnarray}
V(\sqrt{s})_{ij}=-\frac{C_{ij}}{4f^{2}}(\sqrt{s}-M_{i}-M_{j})
\sqrt{\frac{M_{i}+E_{i}}{2M_{i}}}\sqrt{\frac{M_{j}+E_{j}}{2M_{j}}},
\label{omker}  
\end{eqnarray}
where $E_i$ stands for the $i$th baryon energy.     

In order to solve the BS scattering equation, we have to introduce the
regularization to remove the divergence.  We first use the dimensional 
regularization (DIM) as in
Refs.~\cite{Oset:1997it,Oset:2001cn,Nieves:2001wt}.  For convenience  
without loss of generality, the indices representing the coupled
channels will be omitted from now on.  Employing the dimensional
regularization, we obtain the familiar result for the two-body
propagator~\cite{Oller:2000fj}:
\begin{eqnarray}
G(\sqrt{s})_{\rm DIM}=\frac{2M}{16\pi^{2}}\left\{
\frac{m^{2}-M^{2}+s}{2s}\ln\frac{m^{2}}{M^{2}}+\frac{\xi}{2s}
\ln\frac{M^{2}+m^{2}-s-\xi}{M^{2}+m^{2}-s+\xi}\right\}+
\frac{2M}{16\pi^{2}}\ln\frac{M^{2}}{\mu^{2}},  
\label{gdim}  
\end{eqnarray}
where
\begin{equation}
\xi=\sqrt{{(M^{2}-m^{2}-s)}^{2}-4sm^{2}}=
\sqrt{\left(s-(M-m)^{2}\right)\left(s-(M+m)^{2}\right)}\nn  
\end{equation}
related to on-shell center of mass momentum of meson-baryon system,
$\xi/(2\sqrt{s})$. Also it is related to the phase space
($\rho=\frac{M\xi}{4s\pi}$) or the flux 
factor which governs the imaginary part 
of the $T$-matrix. $\mu$ stands for the renormalization scale, which
contains information on the divergence of $G(\sqrt{s})$ with the  minimal 
subtraction.

Now, we introduce form factors instead of the dimensional
regularization.  The form factors encode the complicated extended
structure of hadrons and are usually parameterized in the form of: 
\begin{equation}
F(q^{2})=\left(\frac{\Lambda^{2}-m^{2}}{\Lambda^{2}-q^{2}}\right)^{n},
\end{equation}
where $\Lambda$ denotes a four dimensional cut-off parameter.  In the
case of $n=1$, {\em i.e.} monopole type of the form factor (MF), it is 
equivalent to the well-known {\em Pauli-Villars}
regularization.  The loop integral with the monopole type of the form
factor can be performed analytically: 
\begin{eqnarray}
G(\sqrt{s})_{\rm MF}&=&\int\frac{d^{4}q}{{(2\pi)}^{4}}
\frac{2M}{\{{(P-q)}^{2}-M^{2}\}(q^{2}-m^{2})}
\frac{(\Lambda^{2}-m^{2})}{(\Lambda^{2}-q^{2})}\cr
&=&\frac{2M}{16\pi^{2}}\Big\{ 
\frac{m^{2}-M^{2}+s}{2s}\ln\frac{m^{2}}{M^{2}}+
\frac{\xi}{2s}\ln\frac{M^{2}+m^{2}-s-\xi}{M^{2}+m^{2}-s+\xi}\cr
&+&\frac{(M^{2}-m^{2}-s)}{2s}\ln\frac{\Lambda^{2}}{M^{2}-m^{2}+\Lambda^{2}}
-\ln\frac{M^{2}-m^{2}+\Lambda^{2}}{M^{2}}\cr
&-&
\frac{\eta}{2s}\ln\frac{M^{2}-m^{2}-s+2\Lambda^{2}-\eta}{M^{2}-m^{2}-s+2\Lambda^{2}+\eta}\Big\}, 
\label{gmf}  
\end{eqnarray}
where
\begin{equation}
\eta =\sqrt{{(M^{2}-m^{2}-s)}^{2}-4s\Lambda^{2}}.
\label{eq:eta}  
\end{equation}
Though the $G(\sqrt{s})_{\rm MF}$ shows a similar structure to
$G(\sqrt{s})_{\rm DIM}$ in Eq.~(\ref{gdim}), their analytic behaviors
are rather different.  It is straightforward to calculate all the
cases of $n\geq 1$ by using the following 
recursion formula:
\begin{eqnarray}
G_{n+1}(\sqrt{s})=G_{n}(\sqrt{s})-\frac{\Lambda^{2}-m^{2}}{2n\Lambda}\frac{\partial
}{\partial \Lambda}G_{n}(\sqrt{s}).
\label{grecur}  
\end{eqnarray}
For example, the analytic form of $G(\sqrt{s})_{\rm DF}$ for the
dipole-type form factor ($n=2$) becomes
\begin{eqnarray}
G(\sqrt{s})_{\rm DF}=G(\sqrt{s})_{\rm MF}
-\frac{M(\Lambda^{2}-m^{2})}{8\pi^{2}} \left\{\frac{1}{\eta}
\ln\frac{M^{2}-m^{2}-s+2\Lambda^{2}-\eta}{M^{2}-m^{2}-s+2\Lambda^{2}+\eta} 
\right\}. 
\label{gdf}  
\end{eqnarray}
The divergence can be canceled again by introducing the counter
terms with a  subtraction parameter $a$:  
\begin{equation}
G\rightarrow G+\frac{2M}{16\pi^2}a,
\label{para}  
\end{equation}
In coupled channel calculations, $G$ and $a$ are diagonal elements of
a matrix in the meson-baryon channels.
The subtraction parameter $a$ in Eq.~(\ref{para}) change only the real
part of $G(\sqrt{s})$, while the imaginary part should be
independent of regularization schemes, since it is finite as is
constrained by unitarity.     

\begin{figure}[p]
\begin{tabular}{cc}
\rsz{\includegraphics{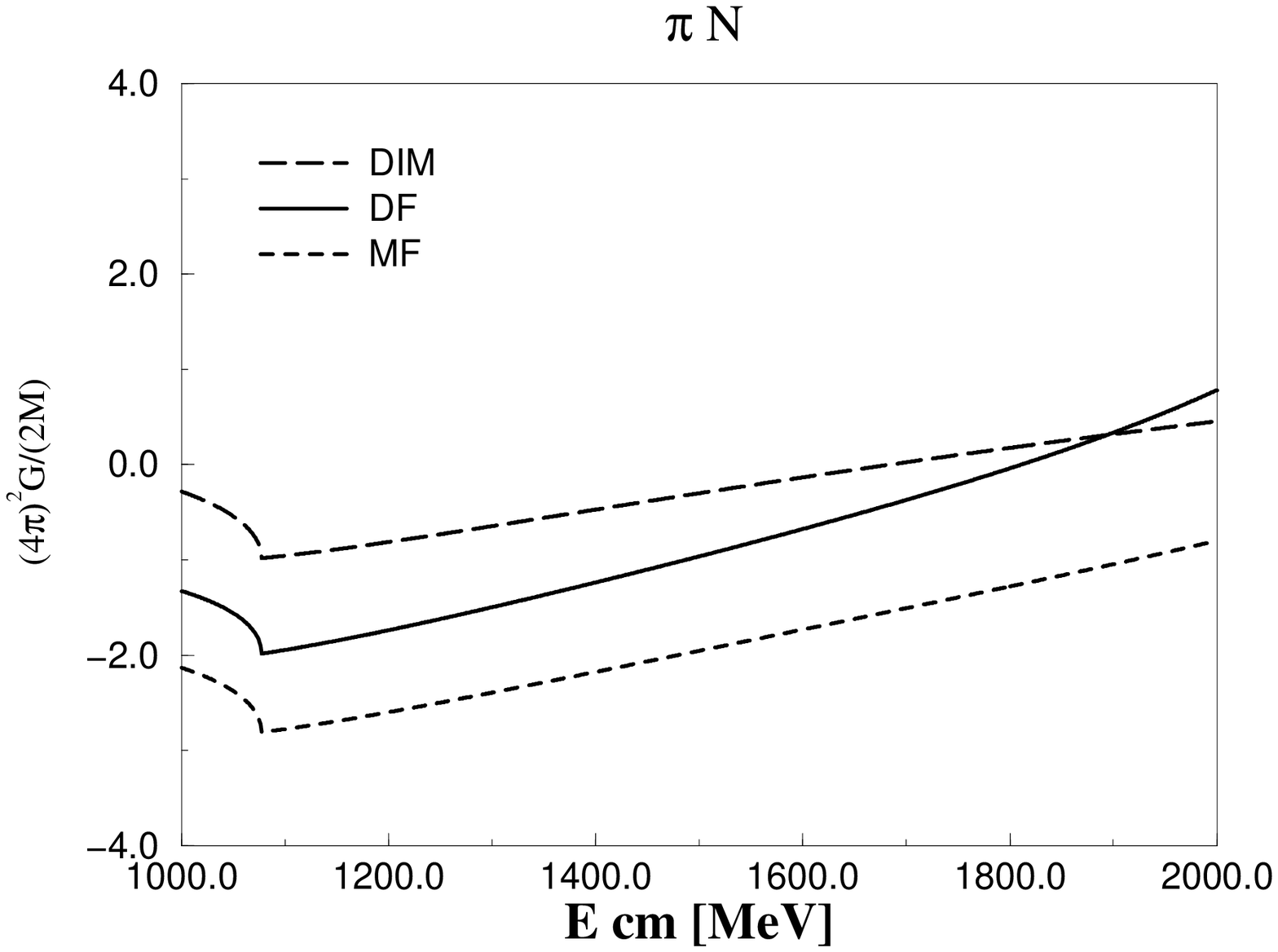}}
\rsz{\includegraphics{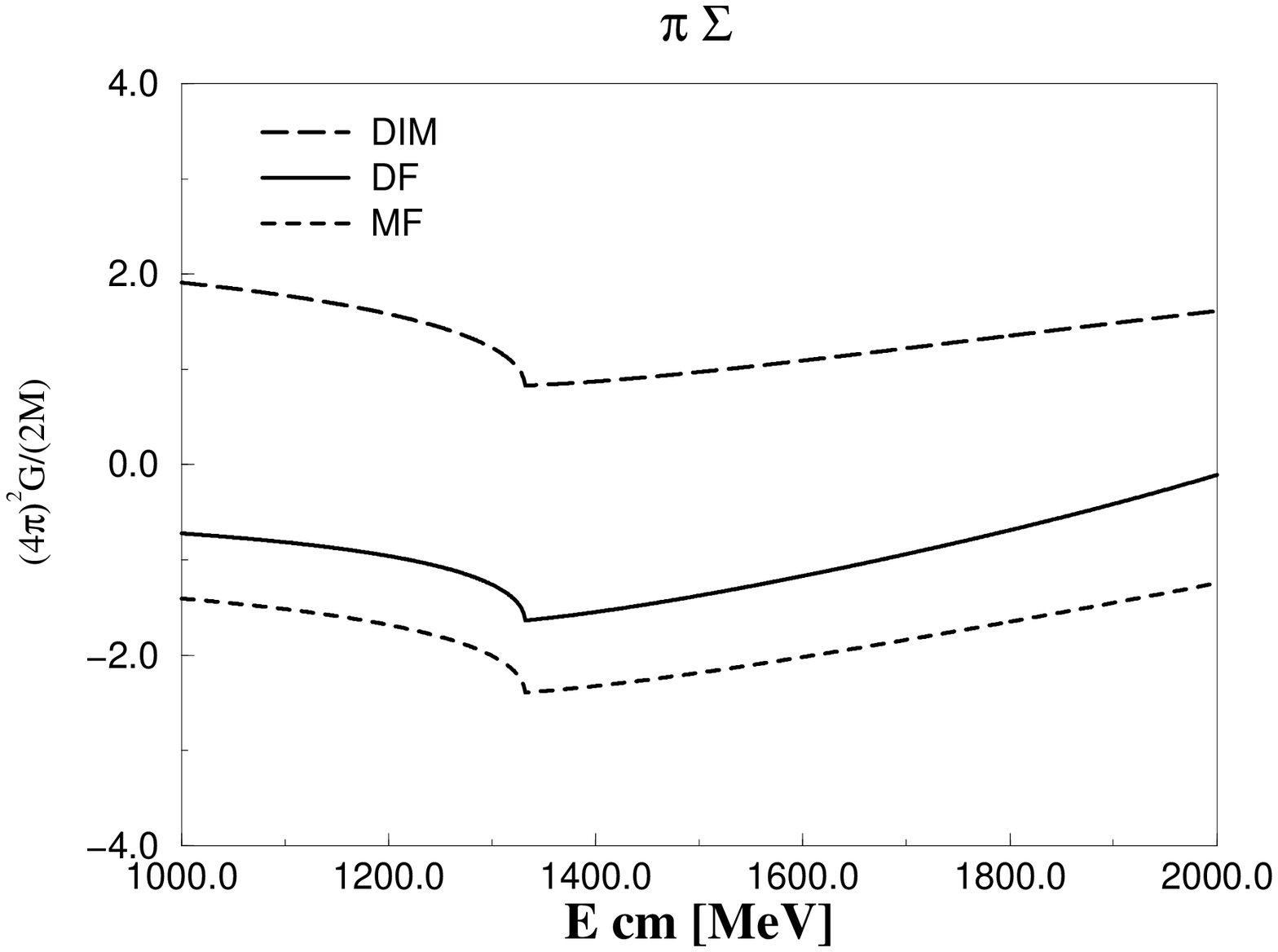}}
\end{tabular}
\caption{The real parts of the meson-baryon loop integrals
for $\pi N$ ($S=-1$,  
$I=0$) and $\pi \Sigma$ ($S=0$,
$I=1/2$) channels as functions of the center of mass energy ${\rm
E_{cm}}(=\sqrt{s})$ for DF (solid), MF 
(dotted) and DIM (dashed).}
\label{g} 
\end{figure}

We plot the real part of $G(\sqrt{s})$ with the DIM, DF, and MF
for four different channels, {\em i.e.} $\pi N$ and $\pi\Sigma$
without the subtraction parameters in 
Fig.~\ref{g}. 
Here, we use $\mu=1200$ MeV for the DIM, whereas   
$\Lambda=1000$ MeV for the DF and MF. Thus, we
can minimize the difference the between regularization schemes by fitting
the subtraction parameters.  However, the results turn out to be
different above the thresholds.  

\begin{figure}[h]
\rsz{\includegraphics{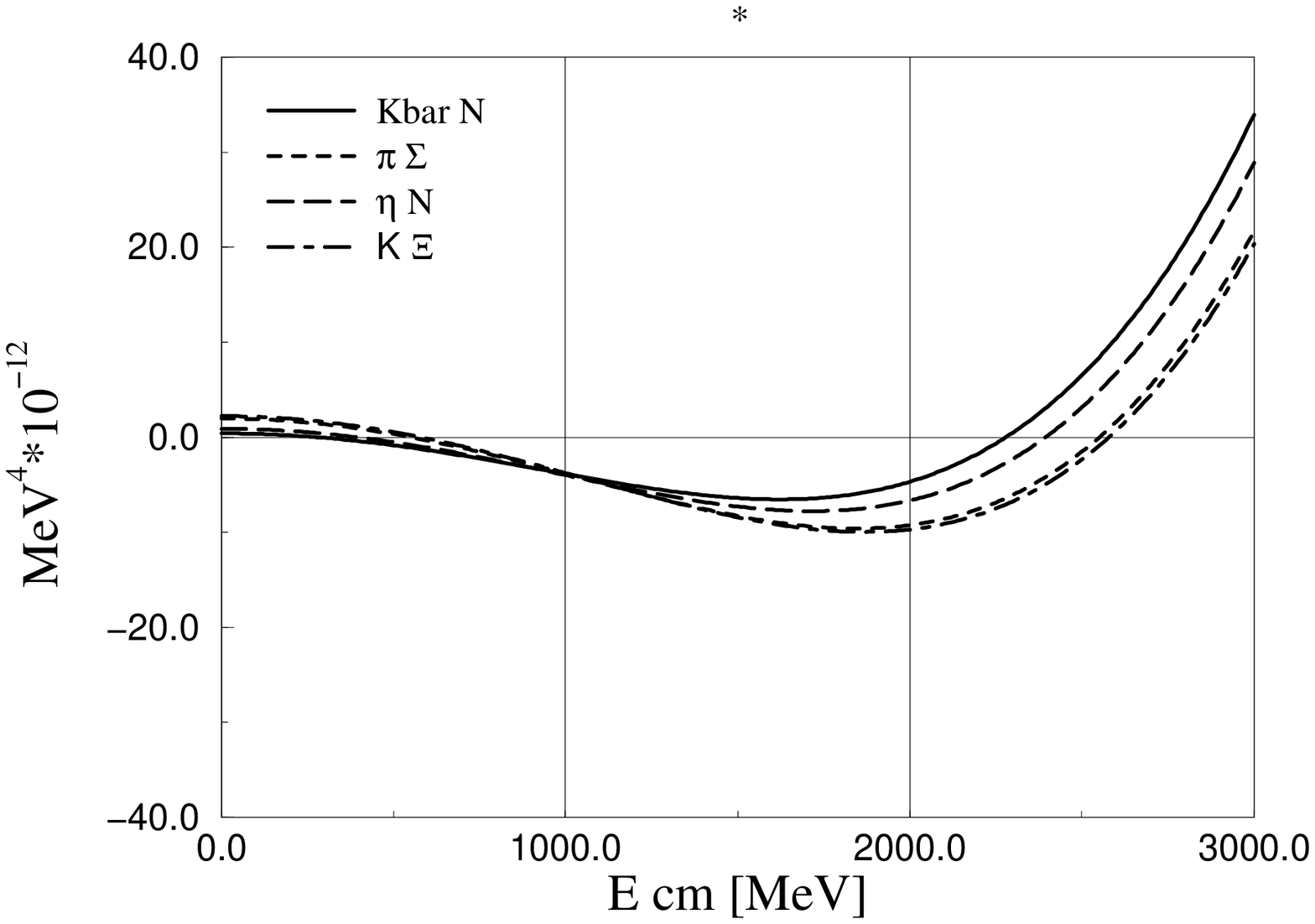}}
\caption[The plots of $\eta^{2} \times 10^{-12}$]
{$\eta^{2} \times 10^{-12}$ as functions of $E_{\rm cm}$ for
$\bar{K}N$ (solid), $\pi \Sigma$ (dotted), $\eta N$ (dashed) and 
$K\Xi$ (dot-dashed) channels.} 
\label{eta}
\end{figure}

In order to find resonances in each channel, we have to look for the
corresponding poles in the second Riemann sheet.  Since the
pseudo-potential is real as given in Eq.~(\ref{omker}), the poles in
the second Riemann sheet are effected only  
by $G(\sqrt{s})$.  The analytic structure of the two-body propagator
is as follows:     
\begin{eqnarray}
G(z)_{\rm 2nd}&=&G(z)_{\rm 1st},\;\;\mbox{for } \sqrt{s}<M+m,\nn \\ 
G(z)_{\rm 2nd}&=&G(z)_{\rm 1st}-2i {\rm Im}G(z)_{\rm
  1st},\;\;\mbox{for }\sqrt{s}\ge M+m,
\label{g12}  
\end{eqnarray}
where $z$ stands for the total energy on the complex plane.  Since the
cut starts from the threshold points, 
the imaginary part of $G(z)$ can be obtained by the discontinuity along
the cut $\sqrt{s}>M+m$:
\begin{equation}
{\rm Im}G(\sqrt{s}) = \frac{\rho}{2} =\frac{M\xi}{8\pi s}
\label{eq:gimag}.  
\end{equation}
In the case of the DIM, we find exactly the same expression for the
${\rm Im} G(\sqrt{s})$ by taking the imaginary part of Eq.(\ref{gdim})
as in Ref.\cite{Inoue:2001ip}.  However, we have two complex
variables, {\em i.e.} $\xi$ and $\eta$ give two different branch
cuts which start from the points of $\sqrt{s}=M+m$ and
where Eq.(\ref{eq:eta}) equals zero in the region of 1 GeV $\sim$ 2 GeV in
the case of the form-factor  
regularizations. Hence, we have to take into account these two variables
to find the poles in the second Riemann sheet.  The variable $\eta$
must be pure imaginary so that Eq.(\ref{eq:gimag}) may be satisfied.
While the on-mass-shell approximation given in Eq.~(\ref{ombs}) works
perfectly well for the DIM, it is to some extent limited for the
form-factor regularization.  However, since we are interested in
resonances in the region of $1\,{\rm GeV} \sim 2\,{\rm GeV}$, in which            
$\eta$ becomes pure imaginary as shown in Fig.~\ref{eta}, the
unitarity condition is also well satisfied in the present case.
Therefore, the term with $\eta$ ($\eta$ term) in 
Eq.(\ref{gmf}) becomes pure real in the region of $1\,{\rm GeV} \sim
2\,{\rm GeV}$. For experimental data of $S=0$ and $S=-1$ we use
Refs.~\cite{Gopal:1976gs, 
Hemingway:1984pz,Hart:1979jx,Saxon:1979xu,Baker:1978bb,Thomas:uh,
Mast:1975pv,Bangerter:1980px,Sakitt:1965kh,Jones:zm,Binford:ts,
VanDyck:ay}.   
\section{Numerical results of ${\rm S}=0$ meson-baryon
sector\label{S0I12}}   
We start with the ${\rm S}=0$ meson-baryon sector from
which $N^{\ast}(1535)$ ($I=1/2$)~\cite{Kaiser:1995cy} and
$\Delta(1620)$ ($I=3/2$)
resonances arise. In 
Table.\ref{s0ch} we list the possible coupled channels.     
\begin{table}[p]
\begin{center}
\caption{\label{s0ch}${\rm S}=0$ channels}
\begin{tabular}{c|cccccc}
 &1  & 2 & 3 & 4 \\
\hline
$I=1/2$&$\pi N $ & $\eta N$ & $K\Lambda$ &
 $K\Sigma$\\
$I=3/2$&$\pi N $ & $K\Sigma$ &-&-\\
\end{tabular}
\end{center}
\end{table}
We fix the parameters to describe the ${\rm S}=0$ sector as follows:
We choose the meson decay constants according to the empirical data
$f_{\pi}=93.0\, {\rm MeV}$, $f_{\eta}=120.9\, {\rm MeV}$, and
$f_{K}=113.46\,{\rm  MeV}$.  The renormalization point for the DIM is
set to be $\mu=1200\,{\rm MeV}$ and the cut-off parameter is fixed to be
$\Lambda=1000$ MeV for the DF and MF.  
We use the subtraction parameters for the DIM, DF and MF as given in
Table.\ref{s0para}. 
\begin{table}[h]
\begin{center}
\caption{\label{s0para}Subtraction parameters for $S=0$ meson-baryon
sector for the DIM,
DF and MF} 
\begin{tabular}{c|cccc}
& $\pi N$ & $\eta N$ & $K\Lambda$ & $K\Sigma$ \\
\hline
{DIM} & 2.0 & 0.2 & 1.6 & -2.8\\
{DF} & 3.00 & 0.93 & 2.38 & -1.87 \\
{MF} & 3.83  & 1.73 & 3.26 & -1.12 \\
\end{tabular}
\end{center}
\end{table}
While we take the subtraction parameters for the DIM from
Ref.~\cite{Inoue:2001ip}, we determine them for the MF and DF in such
a way that the threshold-point values of $G(\sqrt{s})$ are
equal to those of the DIM in order to keep consistency with chiral
perturbation theory, which describes the amplitudes at low energies well. 
 
In order to look into the characteristics of resonances, we need to
investigate the partial-wave amplitudes.   In order to compare the
present results with the experimental data, we normalize the transition amplitudes as follows: 
\begin{eqnarray}
-\sqrt{\frac{M_{i}\xi_{i}}{4\pi\sqrt{s}}}\sqrt{\frac{M_{j}\xi_{j}}{4\pi\sqrt{s}}}  
 T(\sqrt{s})_{ij}. 
\label{pa}  
\end{eqnarray}
Fig.3 represents the partial-wave amplitudes for $\pi N$ scattering in
the
${\rm S}_{11}$ and ${\rm S}_{31}$ channels.  
\begin{figure}[h]
\begin{center}
\begin{tabular}{cc}
\rsz{\includegraphics{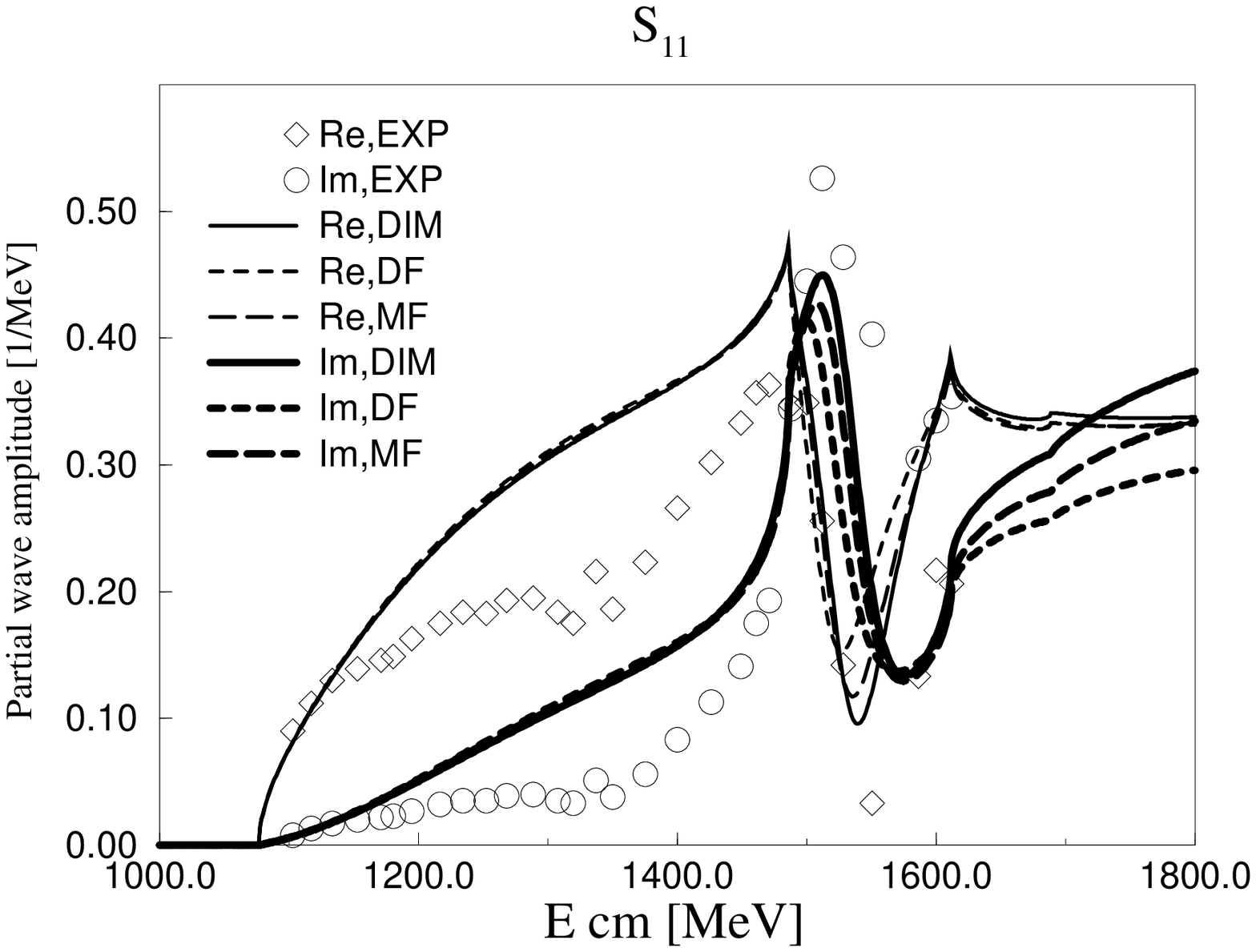}}
\rsz{\includegraphics{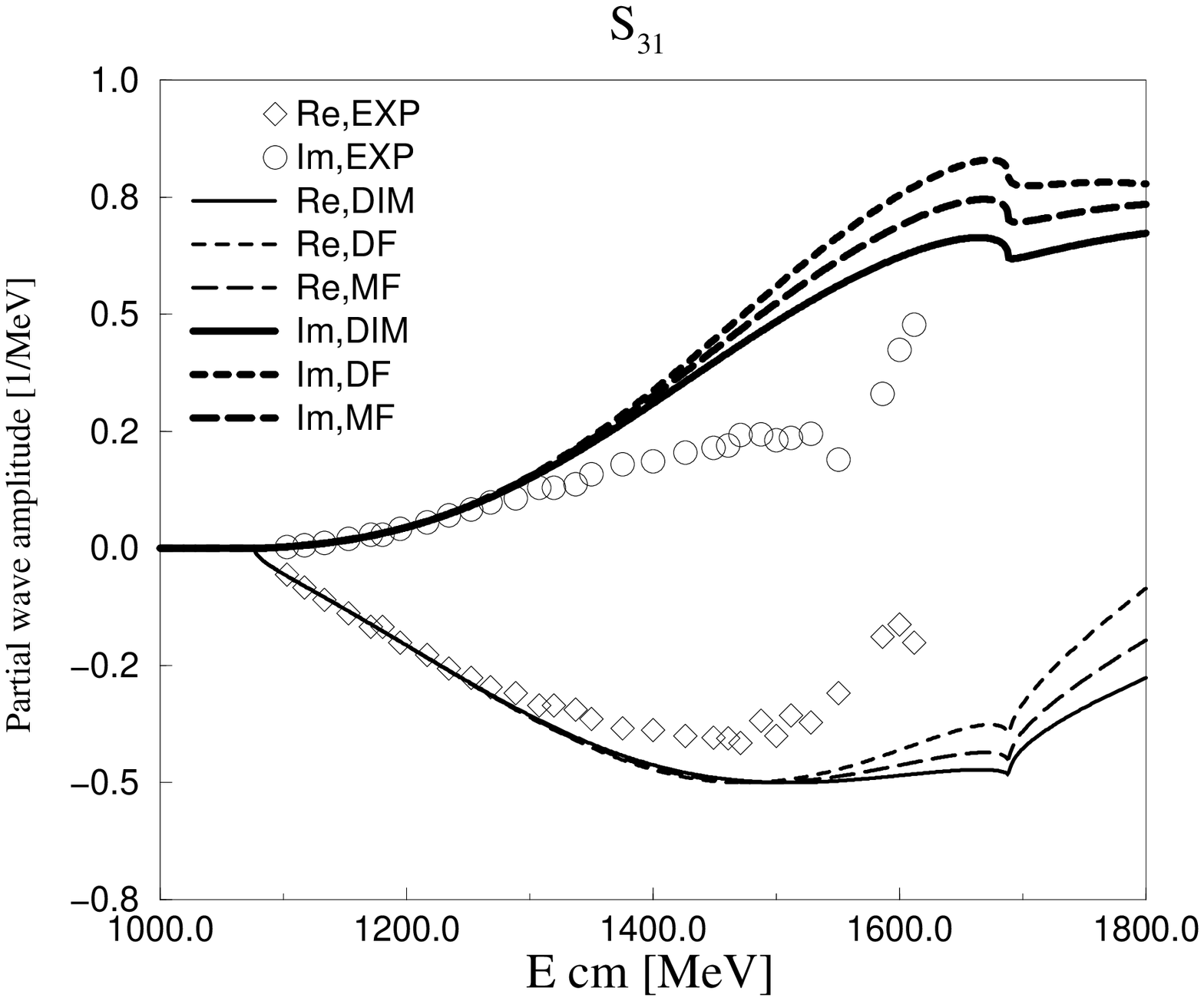}}
\end{tabular}
\caption{$\pi N$ scattering amplitudes ($S=0$) for $S_{11}$ and $S_{31}$
channels as functions of the center of mass energy. Thin and Thick
curves are for real and imaginary parts, where the solid, dotted and
dashed ones for DIM, DF and MF, respectively. Experimental data are
shown by diamonds for the real parts and by circles for the imaginary parts.}
\end{center}
\label{s0pwa}
\end{figure}
We find that the results with the MF and DF are similar to those with
the DIM. The kink around $1500\,{\rm MeV}$ implies the resonance
of $N^{\ast}(1535)$.  Hence, $N^{\ast}(1535)$ is dynamically
generated in the form-factor regularization as well as in the
dimensional one, once the subtraction parameters $a_i$ are properly
chosen.  However, the present calculation fails to reproduce  the higher
resonance $N^{\ast}(1650)$.  Concerning $N^{\ast}(1650)$, we mention
that Ref.~\cite{Nieves:2001wt} has reproduced it by introducing more
parameters than the present work.  Thus, in the present frame work, using
the lowest-order Chiral Lagrangian with the on-mass-shell
approximation is not enough to generate the $N^{\ast}(1650)$ resonance
dynamically.   

\begin{figure}[h]
\begin{tabular}{cc}
\rsz{\includegraphics{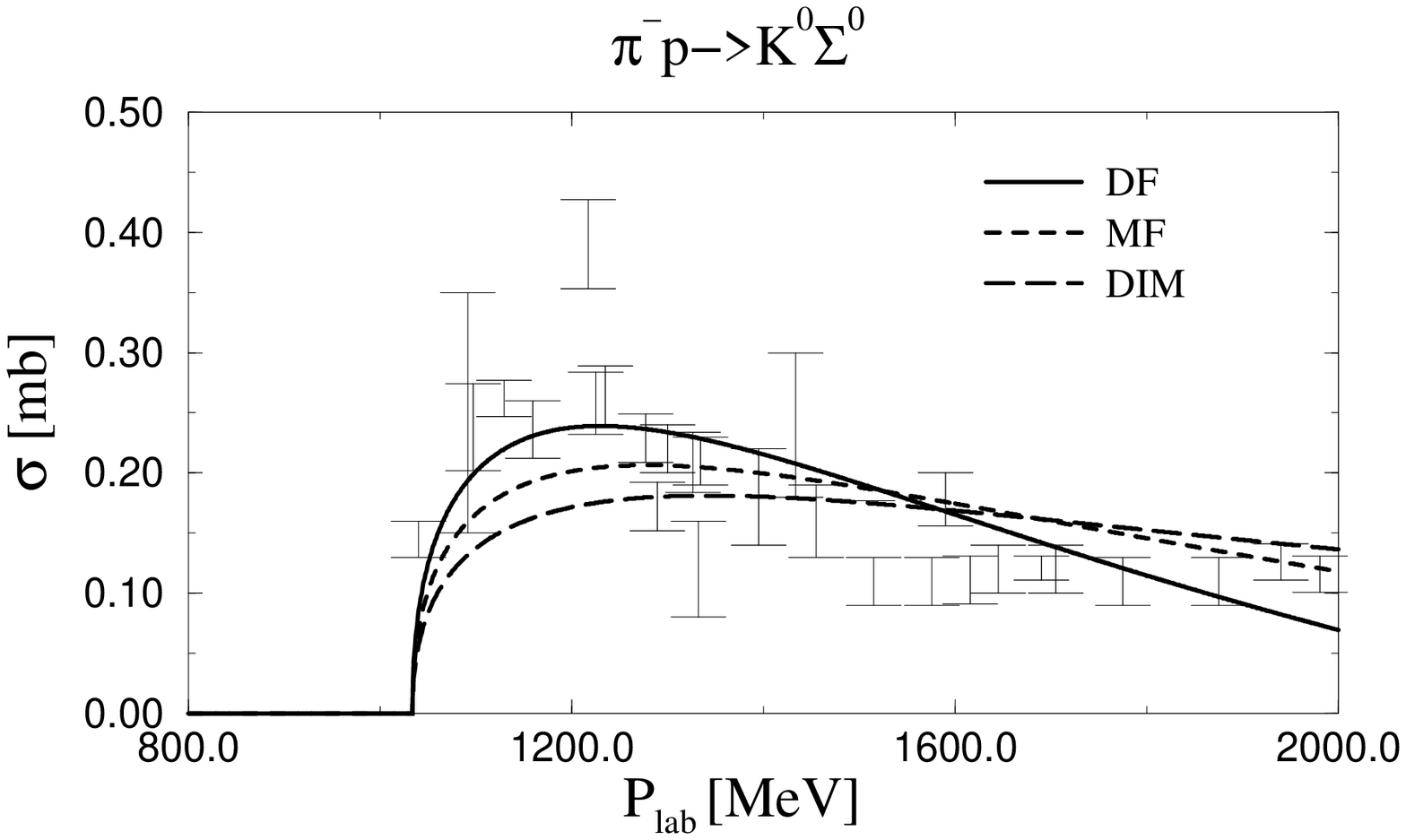}}
\rsz{\includegraphics{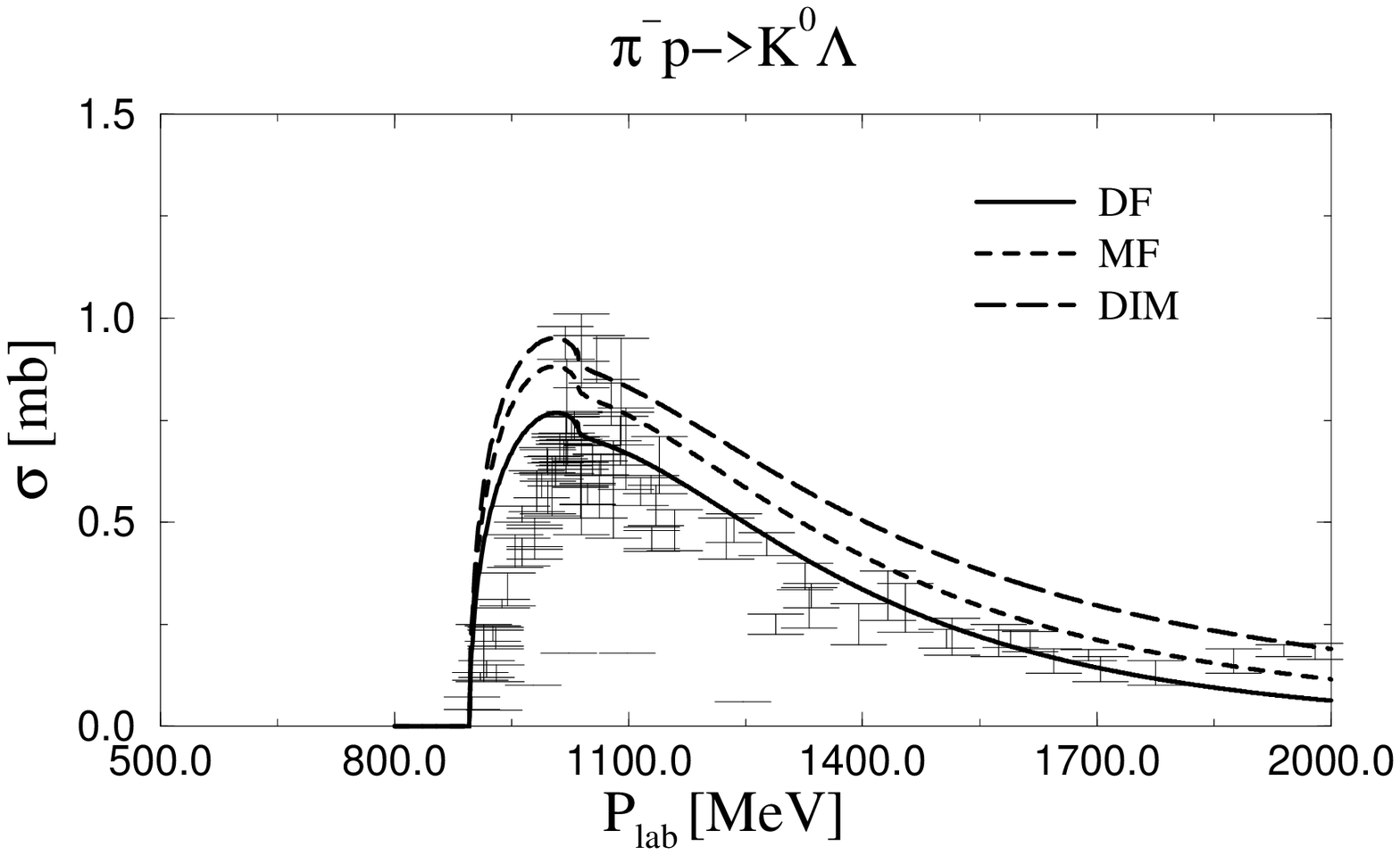}}
\end{tabular}
\begin{tabular}{cc}
\rsz{\includegraphics{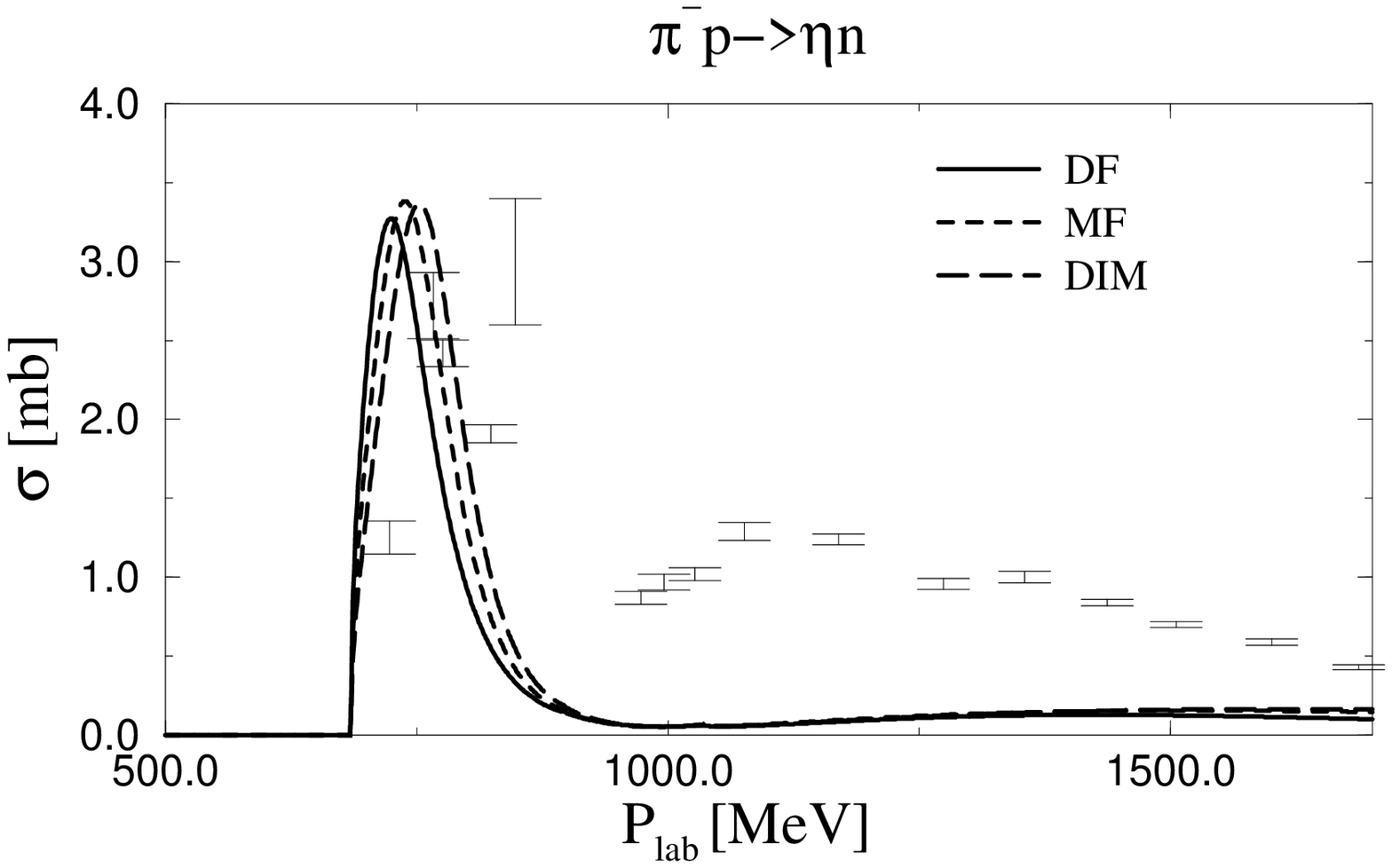}}
\end{tabular}
\caption{${\rm S}=0$, $I_3=-1/2$ cross sections with different
regularizations for $\pi^{-}p\rightarrow K^{0}\Sigma^{0}$, $K^{0}\Lambda$, and $\eta n$ as functions of the
lab momentum $P_{\rm lab}$. The solid curves are for the DF, the dotted one for the MF, and the dashed
one for the DIM, respectively.} 
\label{s0cross}
\end{figure}

In Fig.{\ref{s0cross}}, we draw the total cross sections of as 
functions of the laboratory momentum of $\pi^{-}$.  Both regularization
schemes describe them qualitatively well.  However, there is some
difference in detail, which is due to the fact that the corresponding
propagators shown in Fig.~\ref{g} have different slopes as the
energy increases.  In particular, the total cross section of the
$\pi^- p\rightarrow K^0\Sigma^0$ shows a noticeable dependence on
regularizations.     

Now, we are in a position to look for the poles corresponding to the
resonances in each channel of the ${\rm S}=0$, $I=1/2$ meson-baryon
sector.  In order to find the pole corresponding to the
resonances, we need to extend the $T$ matrix to the complex energy
plane in which the propgator has a cut starting from the
threshold energy.  

We find that the pole is located at $(1516-37i)\,{\rm MeV}$ in the case of the
DIM, while those with the DF and MF are at $(1502-41i)\,{\rm MeV}$
and at $(1517-41i)\,{\rm MeV}$, respectively.  These results are
qualitatively similar to those in Ref.~\cite{Inoue:2001ip}, where
two pions channel and $\rho$ meson exchange are also included.  

We can express an approximated form of the $T$-matrix amplitude
near the pole as follows:
\begin{eqnarray}
T_{ij}\simeq \frac{g_{i}g_{j}}{z-z_{R}},
\label{cs}  
\end{eqnarray}
where the residue $g_{i}$ denotes the coupling strength to the
resonance in the $i$th channel.  Hence, we can easily determine the
coupling strength of the resonance, using Eq.(\ref{cs}).  The coupling 
strengths of $N^{\ast}(1535)$ to each channel of ${\rm S}=0$,$I=1/2$ are
listed in Table~\ref{s0cps}.     
\begin{table}[h]
\begin{center}
\caption{\label{s0cps}Coupling strengths $|g_{i}|$ of $N^{\ast}(1535)$
to four different channels.}
\begin{tabular}{c|cccc}
 & $\pi N$ & $\eta N$ & $K\Lambda$ & $K\Sigma$\\
\hline
{DIM} & 0.46 & 1.79 & 1.09& 5.17\\
{DF} & 0.84 &1.93& 1.60& 3.23 \\
{MF} &1.00 &2.02 &1.78 & 3.54\\
\end{tabular}
\end{center}
\end{table}
We find the following tendency of the coupling strengths:
\begin{equation}
|g_{\pi N}|<|g_{K\Lambda}|\sim|g_{\eta N}|<|g_{K\Sigma}|
\end{equation}
for all regularizations.  With these results we can 
conclude that the $N^{\ast}(1535)$ resonance is strongly coupled to
the $K\Sigma$ channel ~\cite{Kaiser:1995cy}. It is interesting to
compare the above results 
with those in Ref.~\cite{Inoue:2001ip}.  While almost all channels are
similar to Ref.~\cite{Inoue:2001ip}, we get larger value of $|g_i|$
for the $K\Sigma$ channel as quoted in Ref.~\cite{Kaiser:1995cy}.  The
reason lies in 
the fact that the $\pi\pi N$ 
channel may bring down the coupling strength of the $K\Sigma$
channel.   
\section{Numerical results of ${\rm S}=-1$ meson-baryon
sector} 
In this section, we investigate the ${\rm S}=-1$ sector,
emphasizing on the resonances $\Lambda(1405)$ ($I=0$) and
$\Lambda(1670)$  ($I=0$) in
$S$ wave.  The $\Lambda(1405)$ and $\Lambda(1670)$ resonances have
already been studied in Refs.~\cite{Oset:1997it,Oset:2001cn} in the
chiral unitary model in which it was shown that $\Lambda(1405)$
depends weakly on the choice of subtraction parameters while
$\Lambda(1670)$ is very sensitive to them~\cite{Oset:2001cn}.
The fitting procedure  is similar to that in Section~\ref{S0I12}.  We use
$\mu=630\,{\rm MeV}$ for the DIM and $\Lambda=1000\,{\rm MeV}$ for the
DF, and MF.  The average meson decay constant $f=107.18\,{\rm MeV}$ for  
${\rm S}=-1$ is adopted for all regularizations as in 
Ref.~\cite{Oset:1997it}.  The fitted subtraction parameters are given 
in Table~\ref{sm1para}, where those of the DIM are taken from
Ref.~\cite{Oset:2001cn}.  
\begin{table}[h]
\label{sm1para}
\begin{center}
\caption{Subtraction parameters $a$ for the $S=-1$
 meson-baryon sector for the DIM, DF and MF.}
\begin{tabular}{c|cccccc}
& $a_{\bar{K}N}$ & $a_{\pi\Lambda}$ &$a_{\pi\Sigma}$ &
$a_{\eta\Lambda}$ & 
$a_{\eta\Sigma}$ & $a_{K\Xi}$ \\ 
\hline
DIM & -1.84 & -1.83 & -2.00 & -2.25 & -2.38 &
-2.67 \\
DF & 0.22 & 0.59 & 0.47 & -0.12 & -0.20 & -0.36 \\
MF & 1.03 & 1.36 & 1.22 & 0.64 & 0.54 & 0.36 \\
\end{tabular}
\end{center}
\end{table}

\begin{figure}[h]
\begin{tabular}{cc}
\rsz{\includegraphics{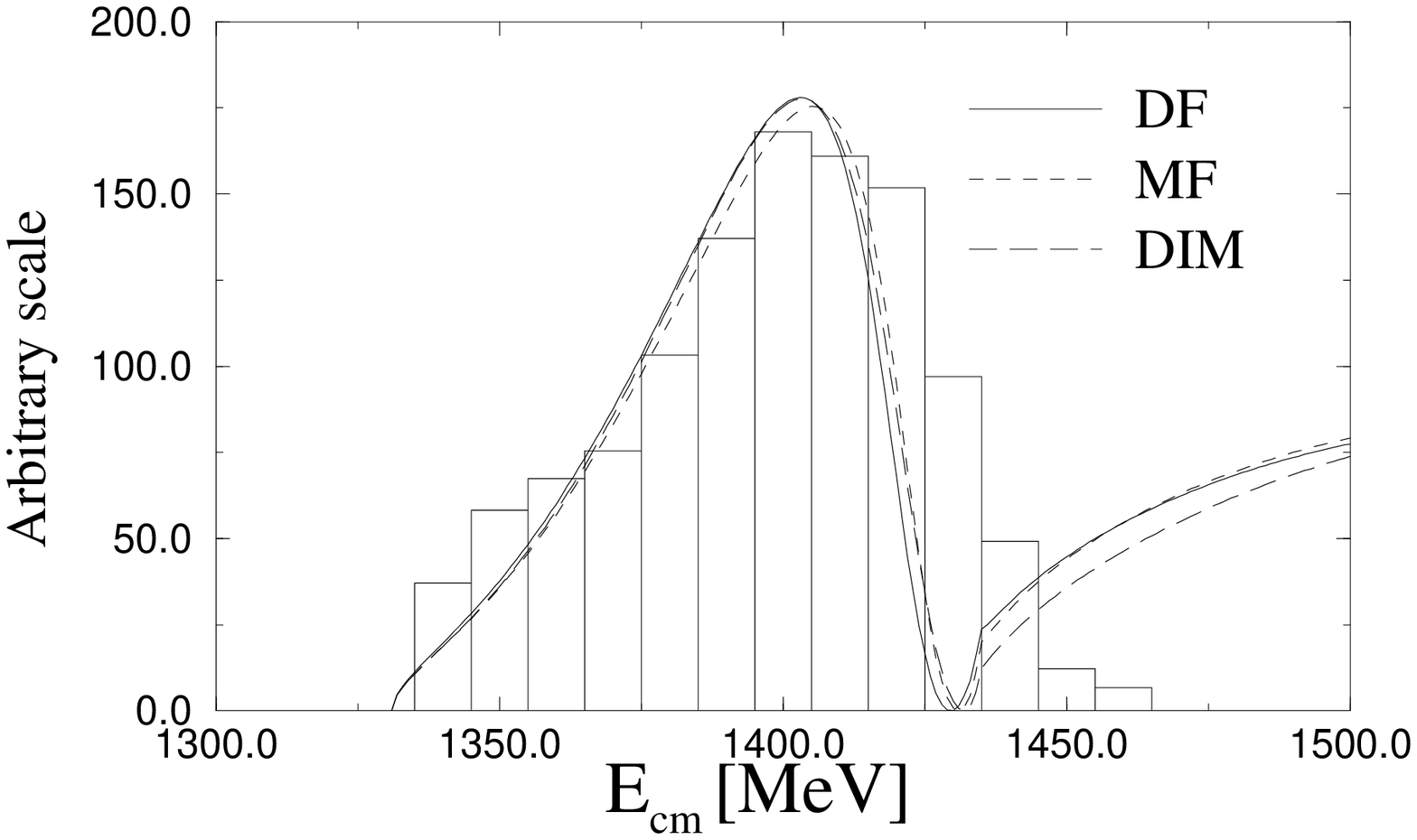}}
\end{tabular}
\caption{$\pi\Sigma$ mass distribution around the $\Lambda(1405)$
  resonance.  The solid curve represents the case of the DM, while the
  dashed and dot-dashed ones correspond to the case of the DIM and MF,
  respectively.  Experimental data is taken from
  Ref.~\cite{Thomas:uh}.}  
\label{sm1mass}
\end{figure}

In the ${\rm S}=-1$, $I=0$ sector it is of great importance to look
into the $\pi\Sigma$ mass distribution around the threshold of
$\bar{K}N$  ($\sim$ 1432 MeV), since the
$\Lambda(1405)$ resonance arises from it. 
Fig.~\ref{sm1mass} shows the $\pi\Sigma$ mass distribution
corresponding to the $\Lambda(1405)$ resonance.  
As shown in Fig.~\ref{sm1mass}, the peak is well reproduced as compared
to the experimental data~\cite{Hemingway:1984pz}.   
There is almost no difference between regularizations.  

\begin{figure}[h]
\begin{tabular}{cc}
\resizebox{8cm}{7cm}{\includegraphics{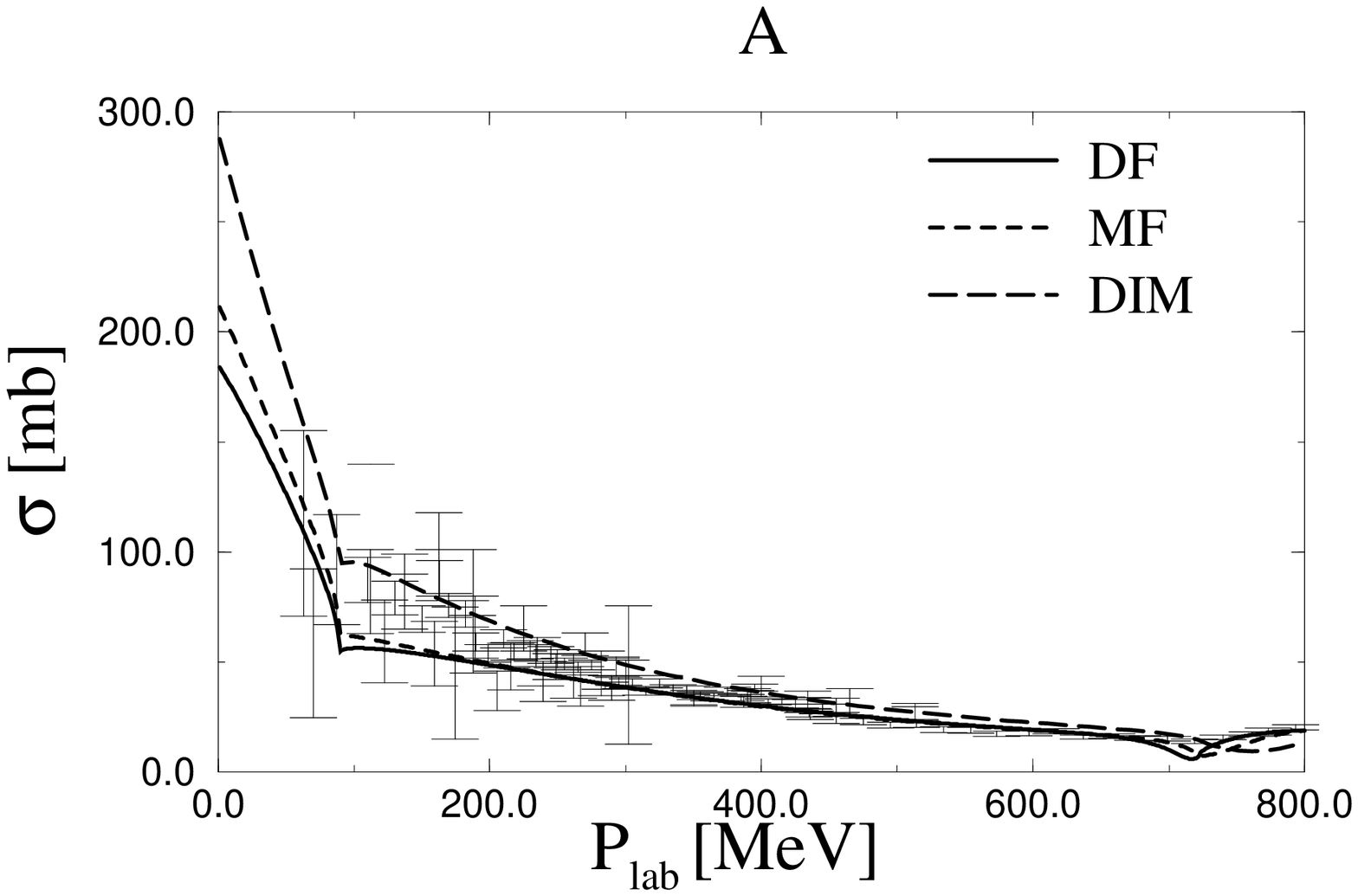}}
\resizebox{8cm}{7cm}{\includegraphics{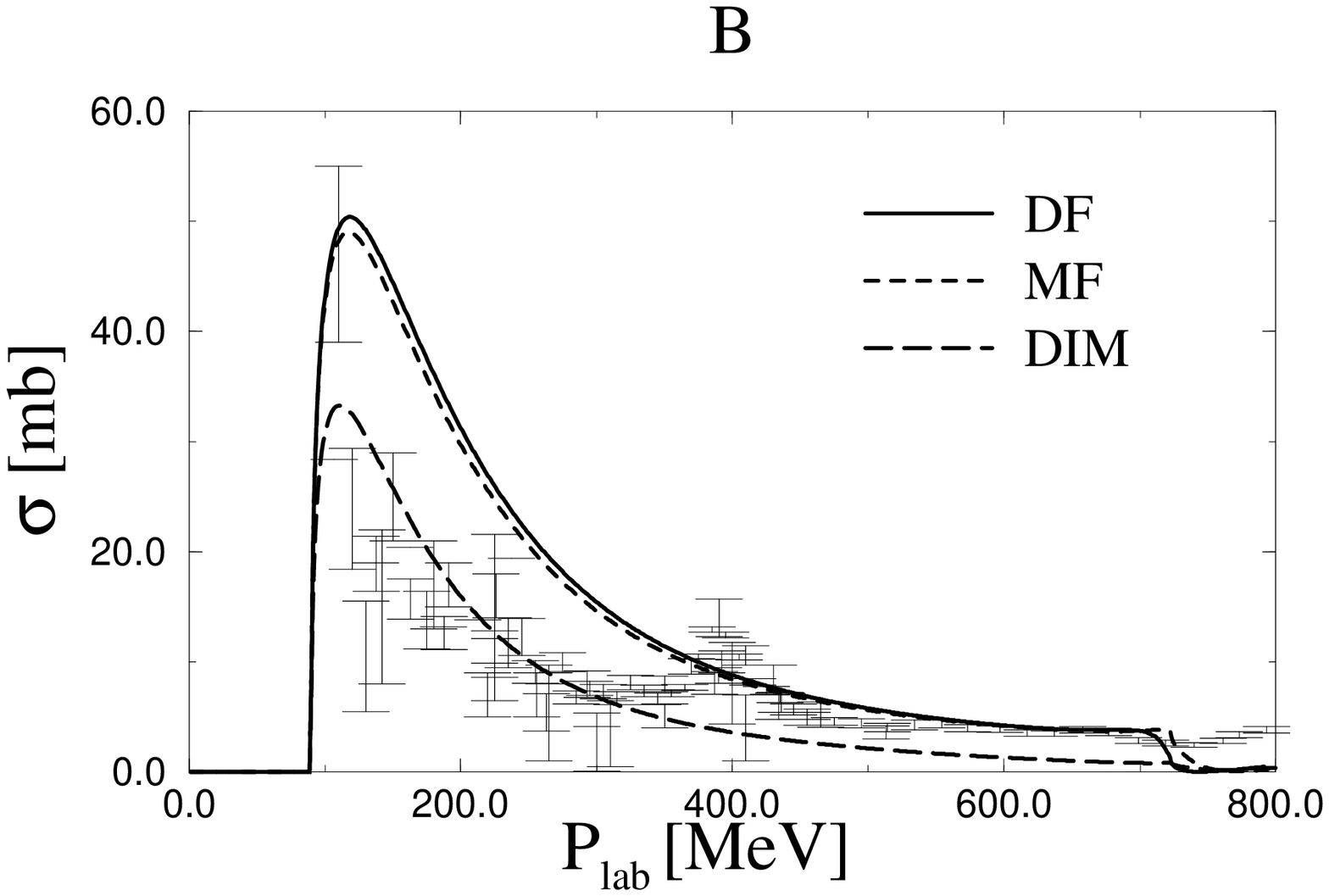}}
\end{tabular}
\begin{tabular}{cc}
\resizebox{8cm}{7cm}{\includegraphics{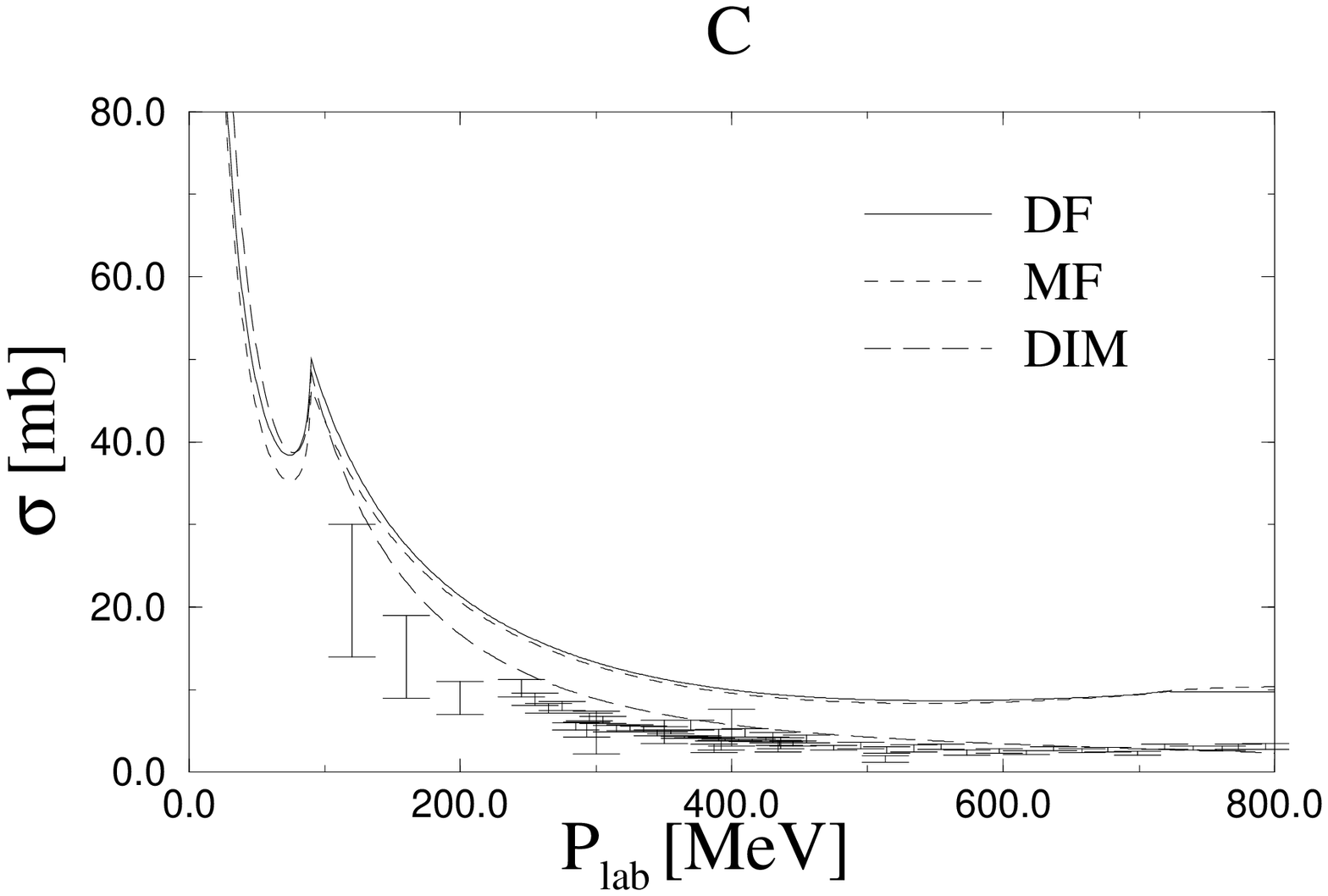}}
\resizebox{8cm}{7cm}{\includegraphics{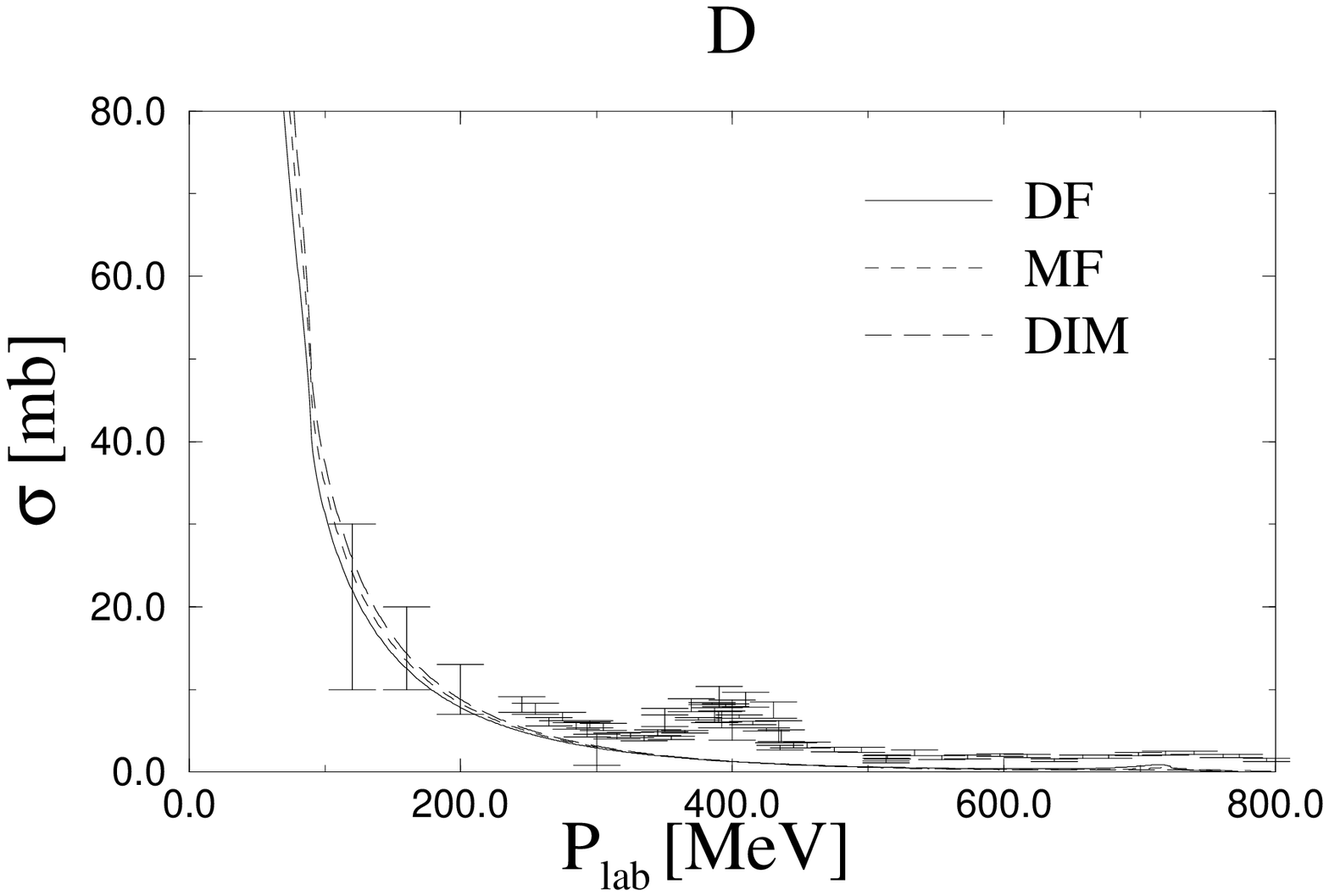}}
\end{tabular}
\begin{tabular}{cc}
\resizebox{8cm}{7cm}{\includegraphics{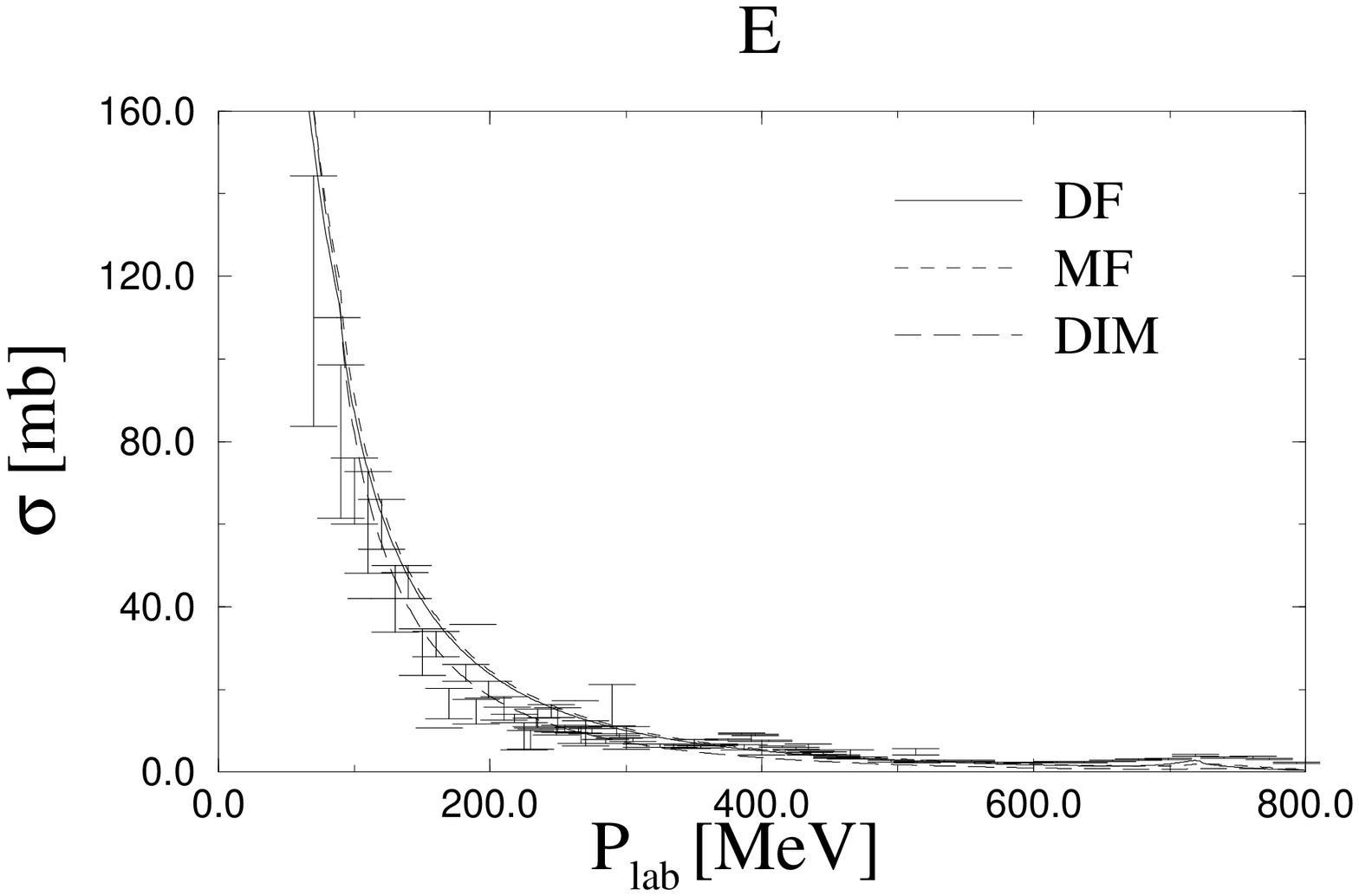}}
\resizebox{8cm}{7cm}{\includegraphics{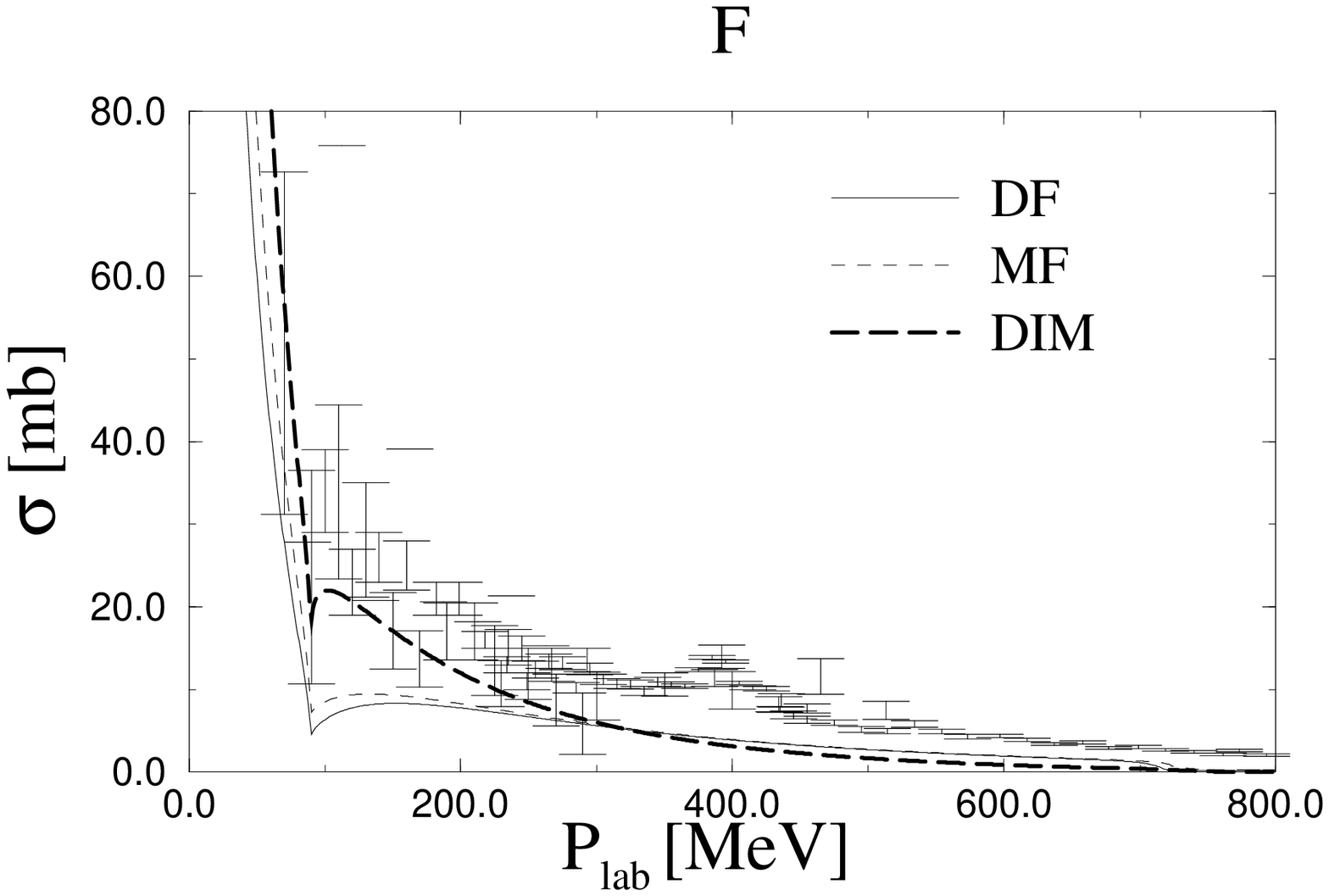}}
\end{tabular}
\caption{$S=-1$,
$I_3=0$ cross sections with different regularizations for  
$K^{-}p\rightarrow K^{-}p$ , $\bar{K}^{0}n$, $\pi^{0}\Lambda$,
$\pi^{0}\Sigma^{0}$, 
$\pi^{+}\Sigma^{-}$, $\pi^{-}\Sigma^{+}$ as functions of the
lab momentum $P_{\rm lab}$. The solid curves are for the DF, the
dotted one for the MF, and the dashed 
one for the DIM, respectively.}
\label{sm1cross}
\end{figure}

The cross sections of the ${\rm S}=-1$, $I_3=0$ are plotted in 
Fig.{\ref{sm1cross}} as functions of the lab momentum for $K^{-}p$
scattering.  The results are in a fairly good agreement with experimental
data.  While difference between regularizations is found in the 
$K^{-}p\rightarrow K^{-}p$, $K^{-}p\rightarrow\bar{K}^{0}n$, and
$K^{-}p\rightarrow \pi^{-}\Sigma^{+}$ processes, there seems no
dependence on the regularization in the case of the $K^{-}p\rightarrow
\bar{K}^{0}n$ and $K^{-}p\rightarrow \pi^{0}\Lambda$ processes.     

We find three different poles in the scattering amplitude in the
$S=-1$, $I=0$ channel. In particular the two poles around $\sim$ 1400
MeV have been confirmed in recent studies based on the chiral unitary
model
~\cite{Hyodo:2002pk,Garcia-Recio:2002td,Jido:2003cb,Hyodo:2003qa,Hyodo:2003jw}.
In 
Table~\ref{poles} and 
Table~\ref{sm1str} we 
list the positions of the poles and coupling strength as done in
Section~\ref{S0I12}.  Three different regularizations give almost the
same results.  
\begin{table}[h]
\begin{center}
\caption{Poles of the scattering amplitude in the $S=-1$ and $I=0$ channel.}
\begin{tabular}{c|ccc}
 & {Pole 1 (MeV)} & {Pole 2 (MeV)} &{Pole 3 (MeV)}\\
\hline
DIM & $1398-74i$ & $1429-14i$ & $1688-22i$\\
DF & $1392-74i$ & $1422-18i$ & $1663-9i$\\
MF & $1394-74i$ & $1425-17i$ & $1671-25i$\\
\end{tabular}
\label{poles}
\end{center}
\end{table}

\begin{table}[h]
\label{sm1str}
\begin{center}
\caption{Coupling strengths $|g_{ii}|$ for three poles.}
\begin{tabular}{c|c|cccc}
& & $\bar{K}N$ & $\pi\Sigma$ & $\eta \Lambda$ & $K\Xi$\\
\hline
    & Pole 1 & 1.43 & 2.06 &0.54 & 0.45\\
DIM & Pole 2 & 2.52&1.30&1.32&0.29 \\
    & Pole 3 & 0.68 &0.16 &0.98 & 3.14\\
\hline
   & Pole 1 & 2.00 & 2.54 &0.78 & 0.62\\
DF & Pole 2 & 3.64 & 1.83 &1.82 &0.53 \\
   & Pole 3 & 0.45 & 0.18 &0.93 & 2.23\\
\hline
   & Pole 1 & 2.65 & 3.57 &1.01& 0.83\\
MF & Pole 2 & 3.39 &1.80 &1.73&0.48 \\
   & Pole 3 & 1.17 &0.36 &1.53 & 4.87\\
\end{tabular}
\end{center}
\end{table}
In the $S=-1$, $I=0$ channel, the tendency of the coupling strengths
are as follows:
\begin{eqnarray}
|g_{K\Xi}|<|g_{\eta\Lambda}|<|g_{\bar{K}N}|<|g_{\pi\Sigma}|
\mbox{   for pole 1}\nn,\\
|g_{K\Xi}|<|g_{\eta\Lambda}|\sim|g_{\pi\Sigma}|<|g_{\bar{K}N}|
\mbox{   for pole 2}\nn,\\
|g_{\pi\Sigma}|<|g_{\bar{K}N}|<|g_{\eta\Lambda}|<|g_{K\Xi}|
\mbox{   for pole 3}\nn.
\end{eqnarray}

\section{Discusstion and Summary}
We have investigated $S=0$ and $S=-1$ meson-baryon scattering,
in particular, focusing on their dependence on regularizations.
Starting from the effective chiral Lagrangian to the lowest order of
the chiral expansion  
$\mathcal{L}^{1}_{MB}$, also known as the Weinberg-Tomozawa term, 
we have constructed the pseudo-potential which is used as a kernel of
the Bethe-Salpeter equation.  In order to solve the Bethe-Salpeter
equation, we utilize the on-mass-shell approximation.  While the
dimensional regularization is widely adopted in almost all works, we
introduced three different schemes of the regularization: The dimensional
regularization, the monopole type form factor (MF), and the dipole type
form factor (DF).  
 
We examined first the dependence of the propagators on
regularizations.  While there is basically no difference among the
regularizations in low-energy regions (below 1500 MeV),  
we found that in higher-energy regions (above 1500 MeV) the propagator
with the dimensional regularization differs substantially from that with
the monopole and dipole-type form factors.  We fixed the subtraction
parameters for the MF and DF at the threshold point.  In addition, we used
empirical values of the branching ratio and the partial wave
amplitudes to fit the $S=-1$ and $S=0$ subtraction parameters for the
form-factor regularizations. For the form factor regularizations, we
have found that the on-mass-shell approximation is limited to the
energy region lower that a certain energy due to the appearance of an
unphysical branch cut. Therefore
further study with the BS scattering 
equation solved numerically is necessary.     

We first calculated the partial-wave amplitudes for the ${\rm S}=0$
meson-baryon sector for $I=1/2$ and $I=3/2$. Though there exists difference
between regularizations in higher-energy regions, the results were
almost the same below 1500 MeV.  The reason lies in the fact that the
propagator in the Bethe-Salpeter equation is rather sensitive to the
regularization schemes in higher-energy region.    
We also calculated the total cross sections for the ${\rm S}=0$
meson-baryon sector.  While all three regularizatoins 
describe well them qualitatively, we found noticeable difference between
regularizations in the total cross section of $\pi^- p\rightarrow
K^0\Sigma^0$.  The position of the resonance $N^\ast(1535)$ is similar
to each other in all regularizaton schemes.  However, the coupling 
strengths of $N^\ast(1535)$ is rather sensitive to regularizations.    

The resonance $\Lambda(1405)$ in the ${\rm S}=-1$ meson-baryon sector
was well reproduced in the present work 
and turned out to be rather insensitive to the types of
regularizations apart from small difference in higher-energy region.
we also investigated the analytic structure of the partial-wave
amplitudes and found that three poles exist in them.  

The regularization dependence of the the total cross sections become
different according to the channels.  While the processes 
$K^-p\rightarrow K^-p$, $K^-p\rightarrow \bar{K}^0 n$, and
$K^-p\rightarrow \pi\Sigma^+$ depend on the  regularization schemes,
even in lower momentum region, those of 
$K^-p\rightarrow \pi^0\Sigma^0$ and $K^-p\rightarrow p\rightarrow
\pi^+\Sigma^-$ show almost no dependence on regularizations.  The
process $K^-p\rightarrow \pi^0\Lambda$ is changed by regularizatons
only in higher momentum region.  The coupling strengths strongly
depend on regularizatons as in the ${\rm S}=-1$, $I=0$ meson-baryon
sector.   

In the present work, we examined the dependence of meson baryon
scattering on regularization schemes.  We employed three different
regularizations: The dimensional regularization, the form factor
regularizations with monopole and dipole types.  The differences due
to regularizatons found in observables are mainly due to the different
behavior of the propagator $G(\sqrt{s})$, because of which in
higher energy region the regularizations change the prediction of the
observables.  Thus, we conclude that one need to vindicate methods
used so far to describe meson-baryon processes.

\section*{Acknowledgments}
HCK is grateful to H. Toki for the warm hospitality at RCNP.  The work
of HCK is supported by the KOSEF (R01--2001--00014). The works of
SINam has been supported by scholarship of the Ministry of
Education, Sciences, 
Sports and Culture of Japan.


\newpage

\begin{thebibliography}{99}
\bibitem{Brown:yv}
G.~E.~Brown, C.~H.~Lee, M.~Rho and V.~Thorsson,
Nucl.\ Phys.\ A {\bf 567} (1994) 937

\bibitem{Lee:1994my}
C.~H.~Lee, H.~Jung, D.~P.~Min and M.~Rho,
Phys.\ Lett.\ B {\bf 326} (1994) 14

\bibitem{Kaiser:1995eg}
N.~Kaiser, P.~B.~Siegel and W.~Weise,
Nucl.\ Phys.\ A {\bf 594} (1995) 325

\bibitem{Krippa:1998ix}
B.~Krippa and J.~T.~Londergan,
Phys.\ Rev.\ C {\bf 58} (1998) 1634 

\bibitem{Nowak} R.J. Nowak {\em et al.}, Nucl. Phys. {\bf B139} (1978)
  61.
\bibitem{Tovee} D.N. Tovee {\em et al.}, Nucl. Phys. {\bf B33} (1971)
493.

\bibitem{Oset:1997it}
E.~Oset and A.~Ramos,
Nucl.\ Phys.\ A {\bf 635} (1998) 99

\bibitem{Oller:2000fj}
J.~A.~Oller and U.~G.~Meissner,
Phys.\ Lett.\ B {\bf 500}, 263 (2001)


\bibitem{Lutz:2001yb}
M.~F.~M.~Lutz and E.~E.~Kolomeitsev,
Nucl.\ Phys.\ A {\bf 700} (2002) 193


\bibitem{Machleidt:hj}
R.~Machleidt, K.~Holinde and C.~Elster,
Phys.\ Rept.\  {\bf 149} (1987) 1.


\bibitem{Mueller-Groeling:cw}
A.~Mueller- Groeling, K.~Holinde and J.~Speth,
Nucl.\ Phys.\ A {\bf 513} (1990) 557.



\bibitem{Meissner:1993ah}
U.~G.~Meissner,
Rept.\ Prog.\ Phys.\  {\bf 56} (1993) 903



\bibitem{Gasser:1984gg}
J.~Gasser and H.~Leutwyler,
Nucl.\ Phys.\ B {\bf 250} (1985) 465.

\bibitem{Inoue:2001ip}
T.~Inoue, E.~Oset and M.~J.~Vicente Vacas,
Phys.\ Rev.\ C {\bf 65}, 035204 (2002)


\bibitem{Oset:2001cn}
E.~Oset, A.~Ramos and C.~Bennhold,
Phys.\ Lett.\ B {\bf 527} (2002) 99
[Erratum-ibid.\ B {\bf 530} (2002) 260]

\bibitem{Nieves:2001wt}
J.~Nieves and E.~Ruiz Arriola,
Phys.\ Rev.\ D {\bf 64} (2001) 116008




\bibitem{Gopal:1976gs}
G.~P.~Gopal, R.~T.~Ross, A.~J.~Van Horn, A.~C.~McPherson, E.~F.~Clayton, T.~C.~Bacon and I.~Butterworth
                  [Rutherford-London Collaboration],
Nucl.\ Phys.\ B {\bf 119} (1977) 362.

\bibitem{Hemingway:1984pz}
R.~J.~Hemingway,
Nucl.\ Phys.\ B {\bf 253} (1985) 742.

\bibitem{Hart:1979jx}
J.~C.~Hart {\it et al.},
Nucl.\ Phys.\ B {\bf 166} (1980) 73.

\bibitem{Saxon:1979xu}
D.~H.~Saxon {\it et al.},
Nucl.\ Phys.\ B {\bf 162} (1980) 522.

\bibitem{Baker:1978bb}
R.~D.~Baker {\it et al.},
Nucl.\ Phys.\ B {\bf 145} (1978) 402; \ Nucl.\ Phys.\ B {\bf 141} (1978) 29.

\bibitem{Thomas:uh}
D.~W.~Thomas, A.~Engler, H.~E.~Fisk and R.~W.~Kraemer,
Nucl.\ Phys.\ B {\bf 56} (1973) 15.

\bibitem{Mast:1975pv}
T.~S.~Mast, M.~Alston-Garnjost, R.~O.~Bangerter, A.~S.~Barbaro-Galtieri, F.~T.~Solmitz and R.~D.~Tripp,
Phys.\ Rev.\ D {\bf 14} (1976) 13.

\bibitem{Bangerter:1980px}
R.~O.~Bangerter, M.~Alston-Garnjost, A.~Barbaro-Galtieri, T.~S.~Mast,
F.~T.~Solmitz and R.~D.~Tripp, 
Phys.\ Rev.\ D {\bf 23} (1981) 1484; 
Phys.\ Rev.\ D {\bf 11} (1975) 3078. 

\bibitem{Sakitt:1965kh}
M.~Sakitt, T.~B.~Day, R.~G.~Glasser, N.~Seeman, J.~H.~Friedman, W.~E.~Humphrey and R.~R.~Ross,
Phys.\ Rev.\  {\bf 139} (1965) B719.

\bibitem{Jones:zm}
J.~J.~Jones {\it et al.},
Phys.\ Rev.\ Lett.\  {\bf 26} (1971) 860.

\bibitem{Binford:ts}
T.~O.~Binford, M.~L.~Good, V.~G.~Lind, D.~Stern, R.~Krauss and E.~Dettman,
Phys.\ Rev.\  {\bf 183} (1969) 1134.

\bibitem{VanDyck:ay}
O.~Van Dyck {\it et al.},
Phys.\ Rev.\ Lett.\  {\bf 23} (1969) 50.

\bibitem{Kaiser:1995cy}
N.~Kaiser, P.~B.~Siegel and W.~Weise,
Phys.\ Lett.\ B {\bf 362}, 23 (1995)

\bibitem{Hyodo:2002pk}
T.~Hyodo, S.~I.~Nam, D.~Jido and A.~Hosaka,
Phys.\ Rev.\ C {\bf 68}, 018201 (2003)

\bibitem{Garcia-Recio:2002td}
C.~Garcia-Recio, J.~Nieves, E.~Ruiz Arriola and M.~J.~Vicente Vacas,
Phys.\ Rev.\ D {\bf 67}, 076009 (2003)

\bibitem{Jido:2003cb}
D.~Jido, J.~A.~Oller, E.~Oset, A.~Ramos and U.~G.~Meissner,
arXiv:nucl-th/0303062.


\bibitem{Hyodo:2003qa}
T.~Hyodo, S.~i.~Nam, D.~Jido and A.~Hosaka,
arXiv:nucl-th/0305011.

\bibitem{Hyodo:2003jw}
T.~Hyodo, A.~Hosaka, E.~Oset, A.~Ramos and M.~J.~Vicente Vacas,
arXiv:nucl-th/0307005.



\end{thebibliography}
\end{document}